# Anisotropic Vapor HF etching of silicon dioxide for Si microstructure release

<u>Vikram Passi</u>[a], Ulf Sodervall[b], Bengt Nilsson[b], Goran Petersson[b], Mats Hagberg[b], Christophe Krzeminski[c], Emmanuel Dubois[c], Bert Du Bois[d], Jean-Pierre Raskin[a]

[a] Université catholique de Louvain (UCL), Information and Communication Technologies, Electronics and Applied Mathematics (ICTEAM), Electrical Department, Maxwell Building, Place du Levant, 3, B-1348, Louvain-la-Neuve, Belgium

[b] Chalmers university of Technology, Department of Microtechnology and Nanoscience - MC2, Nanofabrication Laboratory, SE-41296, Göteborg, Sweden

[c] Institut d'Electronique, de Microélectronique et de Nanotechnologie (IEMN), Silicon Microelectronics Group, Cité Scientifique, Avenue Poincaré, BP 60069, F-59652 Villeneuve d'Ascq Cedex, France

[d] Interuniversity Microelectronics Centre (IMEC), Kapeldreef 75, B-3001, Leuven, Belgium

**Abstract** – Damages are created in a sacrificial layer of silicon dioxide by ion implantation to enhance the etch rate of silicon-dioxide in liquid and vapor phase hydrofluoric acid. The etch rate ratio between implanted and unimplanted silicon dioxide is more than 150 in vapor hydrofluoric acid (VHF). This feature is of interest to greatly reduce the underetch of microelectromechanical systems anchors. Based on the experimentally extracted etch rate of unimplanted and implanted silicon dioxide, the patterning of the sacrificial layer can be predicted by simulation.

## I. Introduction

The two main methods to fabricate microelectromechanical systems (MEMS) are bulk and surface micromachining techniques. In the case of bulk micromachining, fabrication of movable structures is accomplished by selectively etching away the handle substrate underneath the structural layers whereas in surface micromachining series of thin film depositions and selective etching of a particular layer of the stack named the sacrificial layer results in the final desired suspended microstructure. The crucial step to both MEMS fabrication methodologies is the control of the release area and thus the precise definition of the compliant mechanical structures anchors [1] as illustrated in Figures 1a and 1b, showing the underetch of the anchors.

Either wet or dry etching processes can remove the sacrificial layer, where using the former method stiction is encountered, the latter method introduces contamination or residues [2]. Important design considerations for the choice of sacrificial layer are: (i) uniformity and thickness control of the deposited film, (ii) ease of deposition, (iii) etch and deposition rate and (iv) temperature of deposition, as well as (v) etch selectivity. Photoresist has been used as sacrificial layer thanks to its ease to be etched (using oxygen plasma or organic solvents) without harming most of the structural materials [3] [4] [5] [6]. However, the processing is limited to low temperature and hence can be only used with metals as structural layers [1]. A wide variety of MEMS sensors and actuators presented in the literature [7] [8] use polysilicon as structural material with phosphosilicate glass (also called as phosphorous doped glass, PSG) as sacrificial. More recently, poly-silicon-germanium and poly-germanium are being used as structural layers [9]. The great interest of Poly-SiGe is its relatively low deposition temperature of approximately 350°C compared to Poly-Si which is deposited at around 600°C. Poly-SiGe MEMS can then be co-integrated with CMOS integrated circuits in post-process [10]. Besides PSG and low temperature silicon dioxide, thermal silicon dioxide with the increased use of silicon-on-insulator (SOI) substrate as starting material for MEMS applications is used as sacrificial layer.

The release of a microstructure, such as a cantilever presented in Figs. 1a and 1b, is successfully achieved thanks to the isotropic etch of the underlying sacrificial layer. The release of the cantilever of a width $W$ is completed when the sacrificial layer is laterally etched by a length of $W/2$ from both sides of the cantilever. Since the etching is isotropic an underetch of similar length, e.g. $W/2$, appears at the cantilever anchor. On one hand, the isotropic etch characteristic of the etchant (wet or dry) is mandatory to be able to laterally etch away the sacrificial layer underneath the structural layer and thus release it but on the other hand, the isotropic etch of the sacrificial layer is the cause of the anchor underetch. The patterning of the sacrificial layer prior to the deposition of the structural material can be used to overcome the underetch of the anchor region, as shown in Figure 1c for a clamped-clamped beam, but in that case, the structural material presents a step at the anchor regions. Besides the problematic of step coverage for the structural layer, this step raises the complexity of the anchor modeling and therefore the prediction of the mechanical or electromechanical behavior of the MEMS structure.

In this paper, we propose ion implantation into the buried $SiO_2$ sacrificial layer to locally modify the $SiO_2$ etch rate and thus properly define the oxide which must be selectively and isotropically etched away to release the microstructure and the oxide region which must sustain to assure the precise microstructure anchoring. The implanted oxide layer will be etched in either liquid or vapor phase HF.

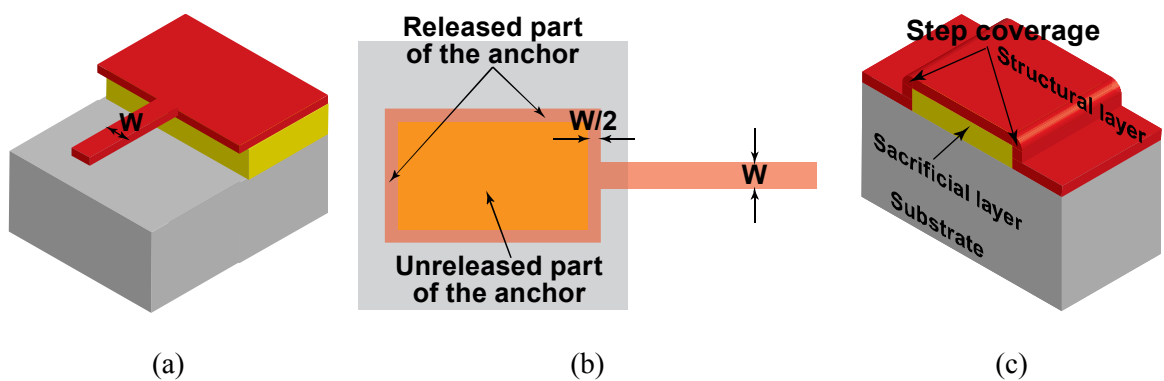

(a)  (b)  (c)

**Figure 1.** (a) Schematic of released cantilever beam, (b) underetch below the anchor pad. The width of the underetch is half the width of the released beam and (c) step coverage of structural material over a predefined sacrificial layer.

**II. VHF etch of implanted silicon dioxide**

Ion implantation is a process by which ions with high energy are introduced into another solid, thus changing the physical and chemical properties of the solid. The introduction of impurities in a semiconductor is the most common application of ion implantation. The three main consequences of ion implantation are: (i) breaking of covalent bonds, (ii) modification of internal stress due to the presence of substitutional atoms, and (iii) modification of semiconducting electrical properties by doping (substitutional atoms). In the case of $SiO_2$, the breaking of covalent bonds between the oxygen and silicon atoms is the dominant phenomenon. The damages created in silicon dioxide enhance the reactivity to hydrofluoric acid (HF) and thus the increase of etch rate [11] - [17]. Selectivity of 3 between implanted and unimplanted silicon dioxide has been demonstrated with the use of wet HF [13] [16] and a selectivity of around 200 when vapor phase hydrofluoric acid is used [16].

Hence, the use of VHF to selectively etch implanted silicon dioxide sacrificial layer regions can be seen as an interesting release methodology where the main advantages such as very high selectivity and dry release are readily exploited. Hereafter, we will present the release of Si cantilevers from a starting SOI wafer where the buried oxide (BOX) is locally implanted to properly define the release area and the anchoring regions (Fig. 2). A germanium (Ge) layer will act as implantation mask to protect the BOX in the anchoring regions. Ge is easily removed by immersing in a preheated solution of hydrogen peroxide at 55°C without affecting any other materials [18]. Once implantation is carried out, the implanted buried silicon-dioxide can be etched selectively with the use of wet HF or preferably VHF as shown in Figure 2b.

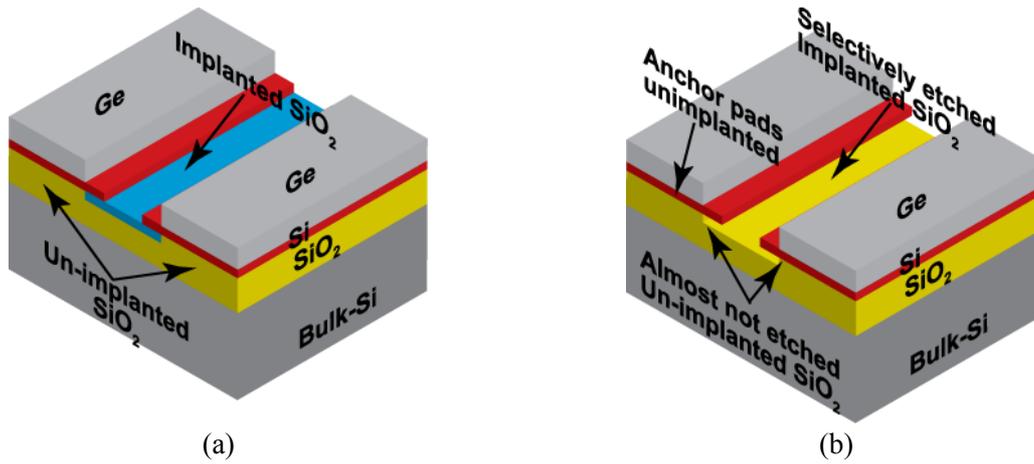

**Figure 2.** Schematics of the test structures: (a) implanted silicon-dioxide (blue region) using Ge layer as an implantation mask, the yellow region shows the unimplanted oxide region, (b) selective etching of the implanted silicon-dioxide in liquid HF solution or VHF. The openings in the germanium are larger than in the top-silicon and the anchor pads in the silicon are unimplanted due to the physical barrier to implantation provided by the germanium.

### III. Samples Fabrication

We would like to release Si microstructures as thin as 20 nm by partially etching away the BOX. Complete process steps are illustrated in Figure 3. The starting substrate is silicon-on-insulator wafer with top-silicon (T-Si) thickness of 160 nm, buried silicon dioxide 400 nm, and 780-µm bulk Si substrate. After cleaning the wafer in mixture of sulphuric acid (90 ml) and hydrogen peroxide (30 ml) followed by HF-1% dip to remove chemical oxide the wafer is subjected to stress free wet thinning of top-silicon using a preheated (68°C) mixture of de-ionized water (1,450 ml), hydrogen peroxide (290 ml) and ammonia (290 ml). Ellipsometer measurement is performed in order to measure the final thickness of top-silicon of 20 nm. Process started by defining marks for alignment using optical lithography. Next, the wafer is spin coated with ZEP-560 positive tone resist at speed of 3,000 rpm for 60 s, baked at 160°C for 60 s to obtain a thickness of 160 nm. Electron beam exposure using proximity correction is performed with JEOL JBX-9300 at 100 keV; a dose of 300 µC/cm² to obtain isolated lines of width from 20 nm up to 2.5 µm. Resist is developed in amylacetate solution for 60 s followed by nitrogen blow drying. Top-silicon is etched using chlorine chemistry using the following parameters: $Cl_2$ - 50 sccm, $Power_{RF}$ – 50 W, $Power_{ICP}$ – 100 W, process pressure – 7 mTorr. Etch stop is performed using a laser interferometer and an over-etch time of 5 s is introduced to ensure complete etching of top-silicon. Resist is removed by immersing the wafer in preheated remover 1165 solution at 65°C for 1 hour. A bilayer resist process is proposed for the lift-off of germanium. Copolymer EL-10% is spin coated at 4,000 rpm, baked at 160°C for 5 minutes to give a thickness of 400 nm. Followed by this ZEP-560 diluted with anisol (1:1) is spin coated at 3,000 rpm, baked at 160°C for 5 minutes to give a thickness of 110 nm. Then, e-beam exposure is done, followed by development in amylacetate solution for 60 s, and nitrogen blow-drying. A 300 nm-thick germanium layer is deposited by evaporation. Lift-off of germanium is performed by immersing the wafer in a bath of preheated remover 1165 solution at 65°C.

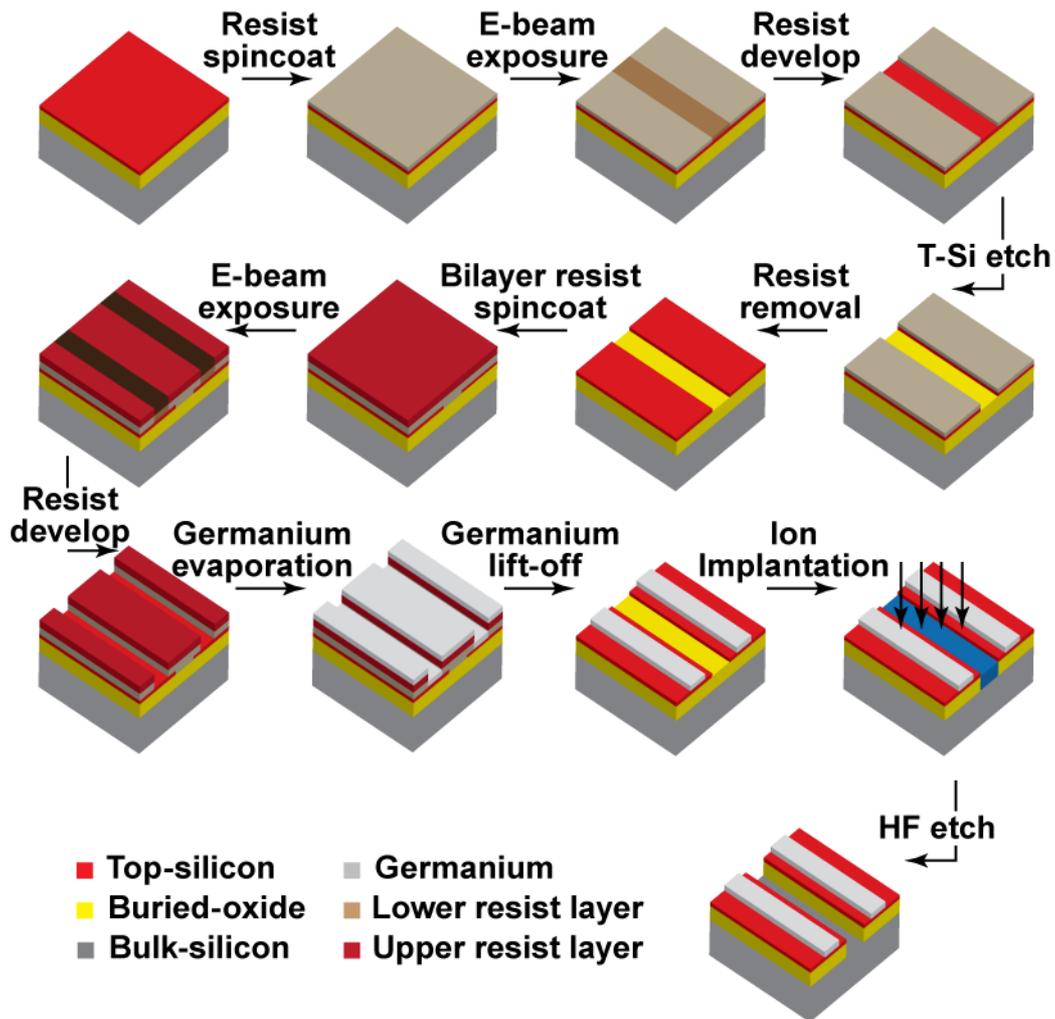

**Figure 3.** Schematic of the complete process flow.

Figures 4a-4b show the top-view SEM images of various openings in silicon and the germanium lift-off, respectively. The area covered by the Ge layer is protected from the species implantation and will correspond to the anchoring regions. After careful inspection of the wafers, the wafers are diced into strips where each strip is implanted with different species. SRIM [19] simulations are performed in order to get the implantation dose and energy parameters. Implantation species used are Arsenic, energy – 110 keV, dose - $7.7 \times 10^{14}$ /cm², Phosphorous, energy – 60 keV, dose – $2.0 \times 10^{15}$ /cm², and Boron, energy – 20 keV, dose – $9.1 \times 10^{15}$ /cm². The implantation is performed at 7° tilt, with the energy tuned in such a way that the implantation depth into the BOX is approximately 100 nm.

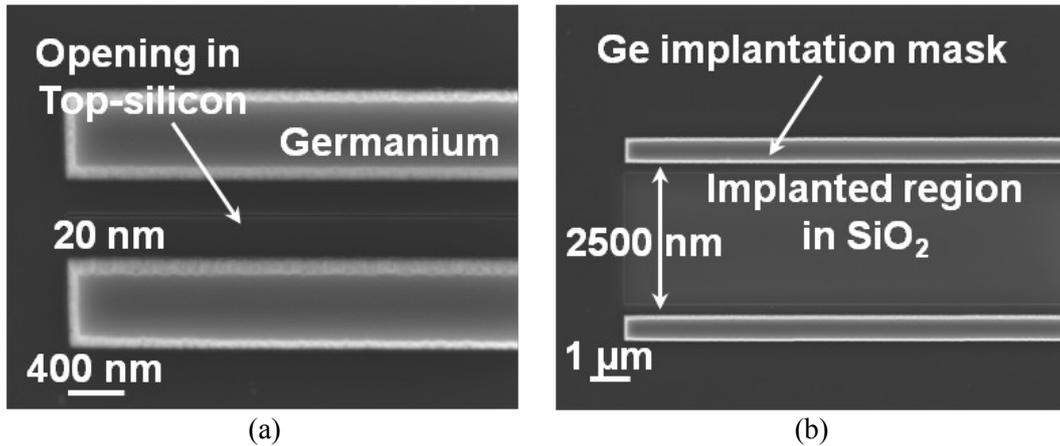

**Figure 4.** Top-view SEM images of Ge implantation mask on top of the patterned thin silicon layer, for the opening of (a) 20 nm and (b) 2,500 nm in the top Si layer. Openings in the germanium layer are regions in the silicon-dioxide which will be implanted.

## IV. Simulation and Experimental Results

The implantation step has been investigated by the mean of Monte-Carlo TCAD process simulations using the Taurus-Process software from Synopsys [20]. Figure 5a presents the 2-D profile of Arsenic in the BOX for the implantation parameters as stated above. Similar profiles have been obtained for P and B implantations. The purpose of the simulation is to estimate the etch profile as function of different process parameters and to compare it with the experimental counterpart. Basically, since the etching rate is expected to be engineered using ion implantation, one-dimensional (1-D) dopant distribution profile and the nuclear deposited energy have been estimated using crystal-trim [21]. Figure 5b presents the nuclear energy profile obtained for the different implantation species. The implantation energy has been defined so that the maximum is located at 100 nm below the silicon-dioxide surface for each species. Next, the etching rate has been calibrated as a function of the nuclear deposited energy using previous experimental results [22]. Subsequently, the variation of the etch rate as a function of the depth is estimated through the variation of deposited energy, and finally integrated as a function of etch time (Figure 5c). Figure 5d shows the nuclear deposited energy as a function of the implanted depth for various implanted species into the silicon-oxide. It can be observed that simulations are in pretty good agreement with the experimental enhanced etching rate in the depth (thickness of the silicon dioxide). It has been shown in [16] that the nuclear deposited energy is the main factor for etch rate modification and the type of implanted species has a limited influence on the etch rate. The etch rate of implanted silicon dioxide does not change in HF-1% when the nuclear deposited energy is lower than $1 \times 10^{23}$ eV/cm³, and there is a saturation of the etch rate for nuclear deposited energy above $3 \times 10^{24}$ eV/cm³. Garrido *et al*. have shown that not more than 15.5% of the silicon-oxygen bonds can be broken [23], and above a certain concentration of nuclear deposited energy a steady state occurs between the breaking and forming of the Si-O bonds.

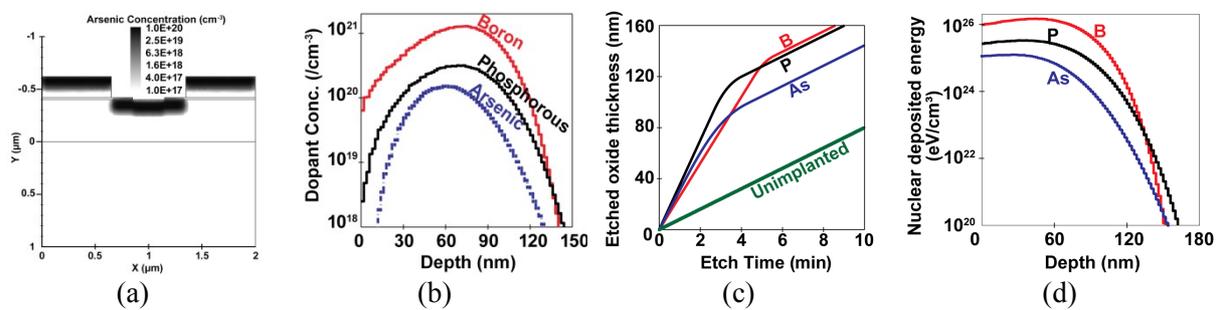

**Figure 5.** Simulation profile of As implantation in silicon dioxide. Energy of implantation used is 110 keV, dose – $7.7 \times 10^{14}$ /cm², tilt – 7°. (a) 2-D implantation profile of As indicating a high concentration over a depth of around 100 nm into silicon dioxide, (b) 1-D implantation profile as function of oxide depth, (c) etched oxide thickness by liquid HF as a function of the time for the different implantation conditions and for the unimplanted case, (d) distribution of nuclear deposited energy for the different implantation conditions.

In the next paragraph, 2-D simulations are compared with the experimental results obtained with liquid HF etching performed for various etch times. Figures 6a-6d show the simulated etch profile of the implanted silicon-dioxide with HF-1% for 2, 5, 10 and 15 mins, respectively and Figures 6e-6h present the cross section SEM micrographs of the implanted silicon-dioxide etched under the same conditions than the simulations. Figures 7a-7f illustrate the simulation and experimental results for the same etch time but for the unimplanted silicon-dioxide, respectively.

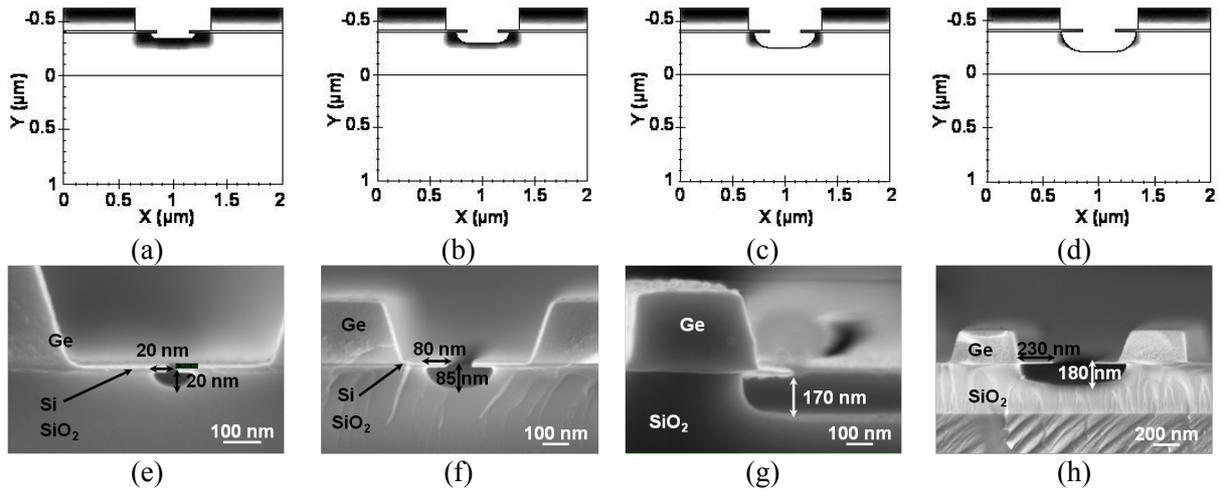

**Figure 6.** Simulated profiles of As implanted silicon dioxide etched with HF-1% for various times: (a) 2 mins, (b) 5 mins, (c) 10 mins and (d) 15 mins. (e)–(h) show the cross section SEM micrographs of the etched silicon-dioxide for the same etching times, respectively.

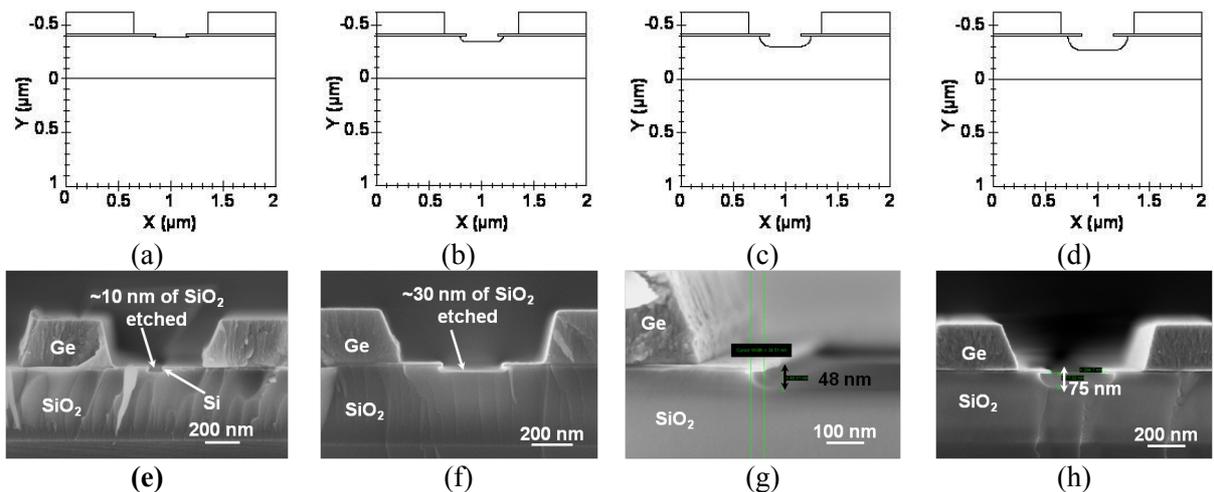

**Figure 7.** Simulated profiles of unimplanted silicon dioxide etched with HF-1% for various time (a) 2 mins, (b) 5 mins, (c) 10 mins and (d) 15 mins. (e)–(h) show the cross section SEM micrographs of the etched silicon-dioxide for the same etching times, respectively.

As can be clearly seen the simulations and the experimental results match very well. The simulation conditions used are very simple although the actual etch mechanism is more complex than what is realized. The simulation does not take into account the amount of water formed during the reaction which causes the etching. This is the reason that the etch depth is slightly larger than the depth indicated by the simulation.

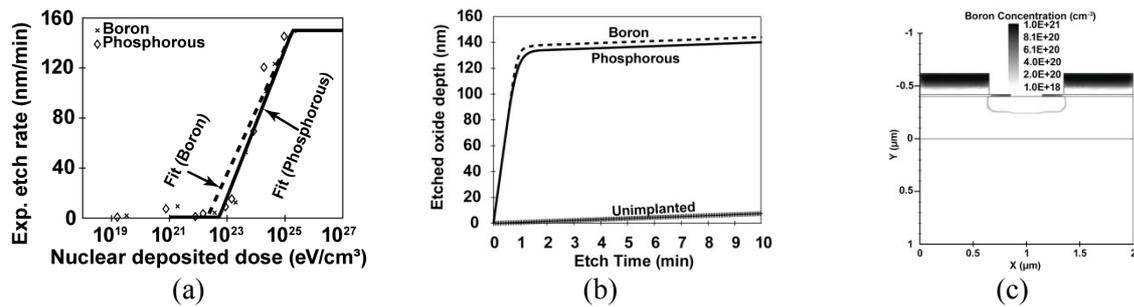

**Figure 8.** (a) Simulation profile nuclear deposited dose as a function of experimental etch rate of implanted silicon-dioxide using VHF for Boron and Phosphorous (Arsenic not shown), (b) Etched oxide depth as a function of etch time. The graph shows clearly the enhancement in etch rate of the implanted silicon-dioxide when compared to the unimplanted silicon-dioxide, (c) 2-D implantation profile of Boron implanted silicon-oxide after VHF etch.

Figure 8 highlights the importance of the implantation of silicon-dioxide and etching with VHF. Figure 8a shows the nuclear deposited dose as a function of the experimental etch rate. As can be seen there is a very narrow window (~$10^{23}$ - ~$3\times10^{24}$ eV/cm³) where there is an enhancement of the etch rate. Below the minimum value the etch rate of silicon-dioxide in VHF is extremely weak (~1 nm/min) and above ~$3\times10^{24}$ eV/cm³ there is a saturation of the etch rate to a value of approximately 150 nm/min. Figure 8b shows the etched silicon-dioxide depth as a function of the etch time. Figure 8b clearly shows that for the unimplanted silicon-dioxide is etched very slowly when compared to the implanted silicon-dioxide. Figure 8c shows the cross section simulation profile of the silicon-dioxide etched with VHF. The cross section shows the clear etching of the silicon-dioxide which is implanted and almost no etching of the silicon-dioxide which is not implanted. This highlights once again the fact that implanted silicon-dioxide when etched with VHF shows better selectivity when compared to etching performed with liquid HF.

Figures 9a-9d and 9e-9h are the cross section profiles of samples etched with VHF at 55°C and 75°C, respectively. In both cases, 5 etch / rinse cycles are carried out with the final rinse at 35°C. As can be clearly seen, the etch depth is larger for etching carried at 55°C. This is due to the slower evaporation of water at 55°C, which is formed during the etching and which acts as a catalyst [16].

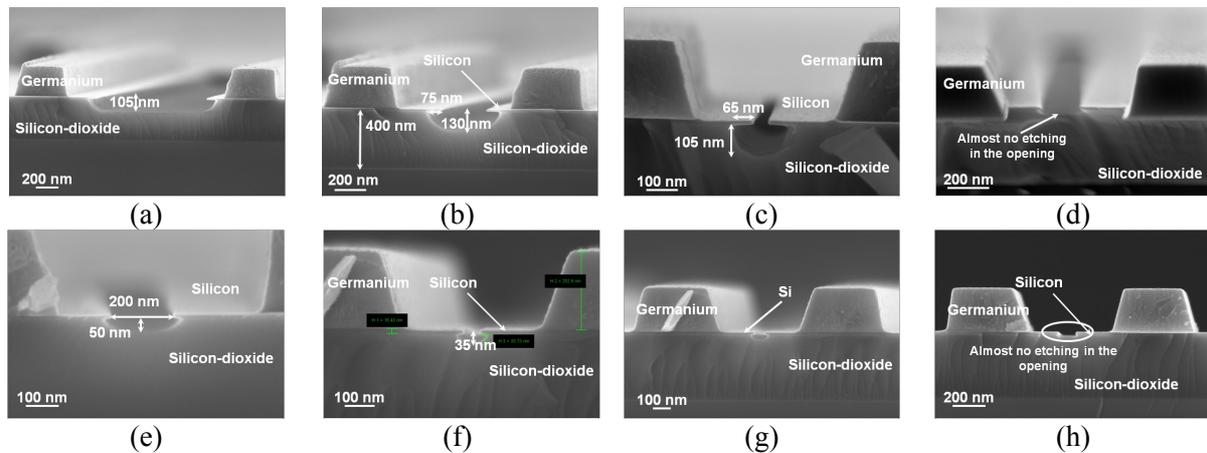

**Figure 9**. Cross-section SEM images after vapor phase hydrofluoric acid (VHF) etching for samples implanted with (a) and (e) Arsenic, (b) and (f) Phosphorous, (c) and (g) Boron and (d) and (h) unimplanted. VHF etching is carried out in a sequence of etch and rinse cycles at 55°C for (a)-(d) and at 75°C for (e)–(h), respectively.

Figure 10 shows the VHF etch process for different etch and rinse cycles. Figure 10a-10b show the etching performed at 55°C for the implanted and unimplanted silicon-dioxide where the etch depth of 135 nm for implanted is obtained when compared to almost no etching for the unimplanted silicon-dioxide (etch rate ratio between implanted and unimplanted oxide of ~150 is observed). A sequence of etch and rinse for 10 mins each is performed. This selectivity is lost or reduced when a longer etch time of 30 mins is used as can be seen in Figure 10c-10d, which is attributed to the fact that for longer etch time the amount of formed water at the sample surface is larger when compared to the shorter etch time (10 mins for Figs. 10a and 10b). In order to maintain a very high selectivity between implanted and unimplanted $SiO_2$, the best process would be to use shorter etch and rinse cycles.

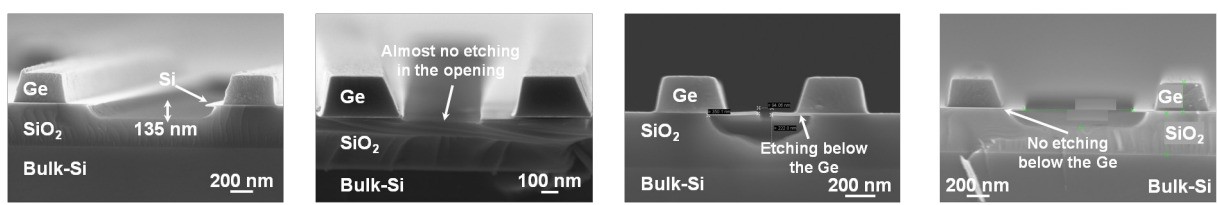

|  (a) | (b) | (c) | (d) |

**Figure 10.** Cross section SEM micrographs of samples etched with VHF at 55°C (a) and (c) implanted SiO$_2$, (b) and (d) unimplanted SiO$_2$. For (a)-(b) etching is performed in a sequence of 5 etch & rinse cycles at 55°C with etch time of 10 mins and rinse time of 10 mins, whereas for (c)-(d) etching is performed with a similar sequence of etch and rinse cycles but with an etch time of 30 mins and rinse time of 10 mins, respectively.

**V. Conclusion**

Implantation of the silicon-dioxide with species such as Arsenic, Phosphorous and Boron is performed to create damages into the silicon-dioxide in order to enhance the etch selectivity to liquid and vapor phase hydrofluoric acid. Etch selectivity of 3 with liquid HF and 150 with VHF are measured between the implanted and unimplanted silicon-dioxide. TCAD simulations of the implantation depth versus dopant concentration and nuclear deposited energy are shown. Both, the implanted and unimplanted oxide are etched using liquid HF-1% for varying time and cross section SEM micrographs are compared with simulation, whereby both show good agreement. Finally, two different sequence of etch and rinse cycles are compared wherein with a shorter etch cycle very high selectivity between implanted and unimplanted oxide is maintained and this selectivity is lost for longer etch time due to the large amount of water formed. The ideal use of the implanted silicon-dioxide etched with VHF is demonstrated by simulation. Implantation of oxide layer can be seen as a technological mean to precisely define buried features such as trenches, cavities, etc. which are revealed using vapor HF. In practice, depending on the thickness of the structural layer and its sensitivity to possible caused damages, the implantation of the silicon dioxide sacrificial layer will be performed before or after the deposition of the structural layer.

**Acknowledgement**

The authors would like to thank Dr. Rémy Charavel from ON Semiconductor for the fruitful discussion and express their gratitude to the MC2, Nanofabrication Laboratory, Chalmers University of Technology, Department of Microtechnology and Nanoscience, Sweden. The authors would especially like to thank the NANOSIL consortium for funding this research and the cleanroom staff at UCL and IEMN for their help in sample preparation.

**References**

[1] B. Bhushan "Springer handbook of Nanotechnology", 2009.

[2] T. Takacs, J. Pulskamp, R. Polcawich, "UV baked/cured photoresist used as sacrificial layer in MEMS fabrication", Army Research Laboratory notes, 2005. http://www.dtic.mil/cgi-bin/GetTRDoc?Location=U2&doc=GetTRDoc.pdf&AD=ADA430096

[3] K. Walsh, J. Norville, Y-C. Tai, "Photoresist as a sacrificial layer by dissolution in acetone", IEEE International conference on Micro Electro Mechanical Systems, 2001, pp. 114-117, MEMS 2001.

[4] I-H, Song, P. K. Ajmera, "Use of a photoresist sacrificial layer with SU-8 electroplating mould in MEMS fabrication", *Journal of Micromechanics and Microengineering*, vol. 13, pp. 816-821, 2003.

[5] S. Soulimane, F. Casset, P-L. Charvet, C. Maeder, M. Aïd, "Planarization of photoresist sacrificial layer for MEMS fabrication", *Microelectronic Engineering*, vol. 84, pp. 1398-1499, 2007.

[6] D. Molinero, L. Castañer, "MEMS switches fabrication suing photoresist as a sacrificial layer", Proceedings of the 2009 Spanish Conference on Electron Devices, pp. 281-284, February 11-13, 2009, Santiage de Compostela, Spain.

[7] S. M. Allameh, P. Shrotriya, A. Butterwick, S. B. Brown, W. O. Soboyejo, "Surface topography evolution and fatigue fracture in polysilicon MEMS structures", *Journal of Microelectromechanical Systems*, vol. 12, no. 3, pp. 313-324, 2003.

[8] A. Ghisi, S. Kalicinski, S. Mariani, I. De Wolf, A. Corigliano, "Polysilicon MEMS accelerometers exposed to shocks: numerical-experimental investigation", *Journal of Micromechanics and Microengineering*, vol.19, pp. 035023, 2009.

[9] S. Sedky, M. Gromova, T. Van der Donck, J-P. Celis, A. Witvrouw, "Characterization of KrF Excimer Laser Annealed PECVD $Si_xGe_{1-x}$ for MEMS Post-Processing", *Sensors and Actuators A: Physical*, vol. 127, no. 2, pp. 316-323, 2006.

[10] A. Witvrouw, "CMOS-MEMS integration today and tomorrow", *Scripta Materialia*, vol. 59, pp. 945-949, 2008.

[11] A. Hirawai, H. Usui, K. Yagi, "Novel Characterization of implant damage in $SiO_2$ by nuclear deposited energy", *Applied Physical Letters*, vol. 54, no. 12, pp. 1106-1109, 1989.

[12] C. Dominguez, B. Garrido, J. Montserrat, J. Morante, J. Samitier, "Etch rate modification in silicon dioxide by ion implantation and rapid thermal annealing", *Nuclear Instruments and Methods in Physics Research Section B: Beam Interactions with Materials and Atoms*, vol. 80-81, no. 2, pp. 1367-1370, 1993.

[13] L. Liu, K. Pey, P. Foo, "HF wet etching of oxide after ion implantation", *IEEE International Electron Devices Meeting*, pp. 17-20, June 1996.

[14] H. Cao, R. J. Weber, "Vapor HF sacrificial etching of phosphorous doped polycrystalline silicon membrane structures", *IEEE International conference on Electro/Information Technology*, pp. 289-293, 2008.

[15] B. Du Bois, G. Vereecke, A.Witrouw, P. De. Moor, C. Van. Hoof, A. De. Caussemaeker, A. Verbist, "HF etching of si-oxides and si-nitrides for surface micromachining", http://www.mrimaxxinc.com/producs/literature/Si-oxides _and_Si-nitrides.pdf.

[16] R. Charavel, J.-P. Raskin, "Etch rate modification of SiO$_2$ by ion damage", *Electrochemical and Solid-State Letters*, vol. 9, no. 7, pp. G245-247, 2006.

[17] P. Holmes, J. Snell, "A vapor etching technique for the photolithography of silicon dioxide", *Microelectronics and Reliability*, vol. 5, pp. 337-341, 1966.

[18] V. Passi, A. Lecestre, C. Krzeminski, G. Larrieu, E. Dubois, J.-P. Raskin, "A single layer hydrogen silsesquioxane (HSQ) based lift-off process for germanium and platinum", *Microelectronic Engineering*, vol. 87, pp. 1872-1878, 2010.

[19] www.srim.org

[20] Synopsys TCAD tools, Taurus Process manuals 2007.03, release 2007.

[21] M. Posselt, B. Schmidt, T. Feudel, N. Strecker, "Atmoistic simulation of ion implantation and its application in Si technology", *Materials Science and Engineering B*, vol. 71, issue 1-3, pp. 128-136, 2000.

[22] R. Charavel, "Etch rate modification by ion implantation of oxide and polysilicon for planar double gate MOS fabrication", PhD Dissertation, Université catholique de Louvain, Louvain-la-neuve, Belgium, 2007.

[23] B. Garrido, J. Samitier, S. Bota, C. Dominguez, J. Montserrat, J. R. Morante, "Structural damage and defects created in SiO$_2$ films by Ar ion implantation", *Journal of Non-Crystalline Solids*, vol. 187, pp. 101-105, 1995.